\documentclass{ws-procs10x7square}

\begin{document}


\title{Fractional Moment Methods for \\ Anderson Localization in the
Continuum\footnote{{\sc\normalsize T}alk delivered by
{\sc\normalsize G}.~{\sc\normalsize S}tolz}}

\author{Michael Aizenman, Alexander Elgart, Sergey Naboko, \\
Jeffrey H.~Schenker and G\"unter Stolz}


\numberwithin{equation}{section}
\newtheorem{thm}{Theorem}
\newtheorem{lem}{Lemma}

\newcommand{\NN}{\mathbb{N}}
\newcommand{\RR}{\mathbb{R}}
\newcommand{\ZZ}{\mathbb{Z}}
\newcommand{\CC}{\mathbb{C}}
\newcommand{\EE}{\mathbb{E}}

\newcommand{\al}{\alpha}
\newcommand{\la}{\lambda}
\newcommand{\ve}{\varepsilon}
\newcommand{\De}{\Delta}
\newcommand{\ga}{\gamma}
\newcommand{\si}{\sigma}

\newcommand{\diag}{\mbox{diag}\,}
\newcommand{\supp}{\mbox{supp}\,}
\newcommand{\dist}{\mbox{dist}\,}
\newcommand{\const}{\mbox{const}\,}
\newcommand{\tr}{\mbox{tr}\,}
\newcommand{\gap}{\mbox{gap}\,}

\maketitle

\abstracts{ The fractional moment method, which was initially
developed in the discrete context for the analysis of the
localization properties of lattice random operators, is extended
to apply to random Schr\"odinger operators in the continuum. One
of the new results for continuum operators are exponentially
decaying bounds for the mean value of transition amplitudes, for
energies throughout the localization regime. An obstacle which up
to now prevented an extension of this method to the continuum is
the lack of a uniform bound on the Lifshitz-Krein spectral shift
associated with the local potential terms. This difficulty is
resolved through an analysis of the resonance-diffusing effects of
the disorder.
 }


\section{Introduction}

The addition of disorder through a random potential may have a
drastic effect on the spectral and dynamical properties of a
Schr\"odinger operator. In certain energy regimes the spectrum of
the operator may turn from absolutely continuous into dense pure
point with localized eigenstates. This phenomenon, known as
Anderson localization, also manifests itself in the form of
dynamical localization, that is the non-spreading of wave packets
supported in the corresponding energy regimes.

There are two known approaches to the mathematical analysis of
localization properties for multidimensional random Schr\"odinger
operators. Both were initially developed in the discrete context,
i.e.\ for random lattice operators. The method of {\it multiscale
analysis} goes back to the ground breaking work of Fr\"ohlich and
Spencer \cite{FS} from 1983 and, by now, has lead to a multitude
of results on spectral and dynamical localization for a wide range
of models. In 1993, Aizenman and Molchanov \cite{AM} introduced
the {\it fractional moment method} into the study of Anderson
localization. For discrete systems this method has provided a
simple perspective on localization and has enabled exponentially
decaying bounds on expectation values of various propagation
kernels.

Multiscale analysis has meanwhile been extended to continuum
Anderson-type models, see e.g.\ \cite{Kotani/Simon, Combes/Hislop}
and, for a state of the art account, \cite{Germinet/Klein}. For an
introduction to multiscale analysis (which is not used in our
work) and many references, see also the recent book
\cite{Stollmann}.

Our goal here is to outline a continuum version of the fractional
moment method and its consequences, with results roughly
corresponding to those obtained for the lattice case in
\cite{ASFH}. We focus on a continuum Anderson-type model in
$L^2(\RR^d)$ of the form
\begin{equation} \label{eq:schrodinger}
H_{\omega} := H_0 + \lambda V_{\omega} \;.
\end{equation}

In Section~2 we will introduce a prototypical set of assumptions
on the background operator $H_0$ and random potential
$V_{\omega}$. We then state three results: First, in Section~3, we
discuss a crucial boundedness result for fractional moments of
``smeared'' Green functions, i.e.\ operator norms of spatially
localized resolvents. Exponentially decaying bounds on this
quantity will then be shown to imply pure point spectrum with
exponentially decaying eigenfunctions as well as exponential decay
for the mean values of transition amplitudes (Section~4). A finite
volume criterion for exponential Green function bounds is provided
in Section~5. Finally, as sketched in Section~6, applications of
the fractional moment method are found by verifying the finite
volume criterion in the usual large disorder, Lifshitz tail or
band edge regimes.

A full account of this work with detailed proofs of all the
results stated below is provided in \cite{AENSS}. For a discussion
of other consequences of the localization results established
here, e.g.\ Kubo conductance and quantum Hall effect, see the
contribution of A.~Elgart to this volume.

\section{A prototypical model}

The operator $H_0$ may incorporate deterministic magnetic and
electric potentials, i.e.\ have the form
\begin{equation} \label{eq:background}
H_0 = (i\nabla -{\bf A}(q))^2 + V_0(q) \;.
\end{equation}
Simple and for our considerations suitable assumptions are local
boundedness of the vector potential $\bf A$, its derivatives
$\partial_i \bf A$ and the positive part $V_{0,+}$ of the electric
potential. We also assume that $V_0$ is bounded from below. Thus
$H_0$ is bounded below and we let $E_0 := \inf \sigma (H_0)$.

Disorder is introduced into (\ref{eq:schrodinger}) through a
parameter $\lambda$ and an Anderson-type random potential
\begin{equation} \label{eq:randpot}
V_{\omega}(q) = \sum_{\alpha\in {\mathcal I}} \eta_{\alpha;\omega}
U_{\alpha}(q) \;.
\end{equation}

For simplicity we will assume here that ${\mathcal I} = \ZZ^d$ and
that $U_{\alpha}(q) = U(q-\alpha)$, $\alpha \in {\mathcal I}$, for
a non-negative, bounded and compactly supported single site
potential $U$, say $\supp U \subset B_0^r$, where $B_x^r =
\{q:|q-x| <r\}$. For technical reasons we also assume that
$|\partial (\supp U)| =0$ ($\partial A$ denoting the boundary of a
set $A$ and $|\cdot|$ $d$-dimensional Lebesgue measure) and that
the $U_{\alpha}$ cover space in the sense that
\begin{equation} \label{eq:cover}
0 < b_- \le \sum_{\alpha \in {\mathcal I}} U_{\alpha}(q) \le b_+ <
\infty
\end{equation}
uniformly in $q\in \RR^d$.

Finally, we assume that the coupling parameters $\eta_{\alpha}$,
$\alpha \in {\mathcal I}$, are independent, identically
distributed random variables with absolutely continuous
distribution $d\mu(\eta) = \rho(\eta)d\eta$. The density $\rho$ is
bounded and supported in $[0,1]$.

A number of these assumptions can be weakened, in particular those
on $\mathcal I$, $U_{\alpha}$ and $\eta_{\alpha}$. The
coefficients of the background operator $H_0$ may include certain
$L^p$-type singularities. For more discussion on what is
technically necessary see \cite{AENSS}.

\section{Finiteness of fractional moments}

A central object in the fractional moment approach to localization
for lattice operators is given by the fractional moments $\EE
(|G_{E+i0}(x,y)|^s)$, where $G_z(x,y) = \langle \delta_x |
(H-z)^{-1} | \delta_y \rangle$ is the Green function, $0<s<1$, and
$\EE$ denotes averaging over the disorder. Finiteness of the
fractional moments is seen relatively easily for suitable
distribution of the random parameters as the singularities of the
Green function become integrable through the exponent $s<1$.

As noted previously in the multiscale analysis approach, for
continuum models a useful counterpart of the discrete Green
function $|G_{E+i0}(x,y)|$ need not be the integral kernel of
$(H-E-i0)^{-1}$, but rather the operator norm $\|\chi_x
(H-E-i0)^{-1} \chi_y \|$ for suitable compactly supported
functions $\chi_x$ and $\chi_y$. Finiteness of their fractional
moments is a crucial preliminary result for our discussion and
technically deeper than the corresponding result in the discrete
case.

To state this result, for an open set $\Omega \subset \RR^d$ we
denote by $H^{(\Omega)}$ the restriction of $H=H_{\omega}$ to
$L^2(\Omega)$ with Dirichlet boundary conditions. Throughout we
also denote characteristic functions of a set $\Lambda$ by
$1_{\Lambda}$ and, for $x\in \RR^d$, $\chi_x = 1_{B_x^r}$, where
the size $r$ of the bumps $U_{\alpha}$ serves as a convenient
length scale.

\begin{lem} \label{lemma}
Let $H_{\omega}$ be a random Schr\"odinger operator as in
(\ref{eq:schrodinger}) with assumptions as in Section~2. Then, for
each $s\in (0,1)$ and $\lambda >0$, there exists $C_{s,\lambda}
<\infty$ such that
\begin{equation} \label{eq:boundedness}
\sup_{\varepsilon >0} \EE \left(\| \chi_x
\frac{1}{H^{(\Omega)}-E-i\varepsilon} \chi_y \|^s \right) \le
C_{s,\lambda} (1+1/\lambda)^s (1+|E-E_0|)^{s(d+2)}
\end{equation}
for any open $\Omega \subset \RR^d$, $x,y \in \Omega$ and $E\in
\RR$. One can choose
\begin{equation} \label{eq:Cbound}
C_{s,\lambda} \le \mbox{const.} \frac{(1+\lambda)^{s(d+2)}}{1-s}
\;.
\end{equation}
\end{lem}

As in the discrete case, the proof of this result proceeds by
showing that the independent variation of some of the random
parameters $\eta_{\alpha}$ resolves singularities which are due to
the proximity of the given energy to an eigenvalue whose
eigenvector has significant support nearby. However, a change in a
parameter can also have the opposite effect, through the creation
of a resonance. In the discrete setup the latter possibility
occurs at not more than a single value of the random parameter,
since each coefficient affects a rank-one term and the number of
energy levels which can be moved past $E$ is bounded by the rank
of the perturbation. Aside from the fact that the rank-one
analysis is not applicable, the source of the difficulty in the
extension of the previous analysis can be traced to the fact that,
in the continuum setup, there is no uniform bound on the
corresponding ``spectral shift''. To circumvent these difficulties
we employ the Birman-Schwinger principle in place of rank-one
analysis, and control the Lebesgue measure of the nearly-singular
values of a coupling parameter by means of the following ``weak
1-1'' type bound
\begin{equation} \label{eq:weakone}
\left| \{ \eta: \|T(\eta+A+i0)^{-1} T\|_{HS} >t \} \right| \le
\frac{C}{t} \|T\|_{HS}^2\;,
\end{equation}
valid for any maximally dissipative operator $A$ and
Hilbert-Schmidt operator $T$. This result was proven in
\cite{Naboko}.

As a consequence of this analysis we find that it suffices to
average over ``local environments'' of $x$ and $y$. Rather than
taking the full expectation one merely averages over the
$\eta_{\alpha}$ with $\alpha$ in suitable neighborhoods of $x$ and
$y$. This yields a bound as in (\ref{eq:boundedness}) with
constants which are uniform in the values of the remaining random
parameters, an improvement of Lemma~\ref{lemma} which is important
in the proof of Theorem~\ref{thm2} below.

\section{Localization properties}

The uniform fractional moment bound (\ref{eq:boundedness}) holds
for all energies in the class of continuum Anderson models
considered here. In the following we will identify the existence
of exponentially decaying bounds (in $|x-y|$) for the left hand
side of (\ref{eq:boundedness}) as a characteristic of the
localization regime. We first show that such bounds for finite
volume operators (but uniformly in the volume) imply spectral and
dynamical localization.

Let $\Omega \subset \RR^d$ be open and $\Lambda_n \subset \Omega$,
$n\in\NN$, a sequence of bounded open domains such that $\bigcup
\Lambda_n =\Omega$ and $H^{(\Lambda_n)}$ converges to
$H^{(\Omega)}$ in strong resolvent sense. We also define a
modified distance by
\begin{equation} \label{eq:distance}
\dist_{\Omega}(x,y) := \min \{ |x-y|, \dist(x,\Omega^c) +
\dist(y,\Omega^c) \} \;.
\end{equation}
$P_{\mathcal J}(H)$ denotes the spectral projection onto $\mathcal
J$ for $H$ and $\|\cdot\|_{\tr}$ the trace norm.

\begin{thm} \label{thm1}
Let $H$, $\Omega$ and $\Lambda_n$, $n\in \NN$, be as above.
Suppose that for some $0<s<1$ and an open bounded interval
$\mathcal J$ there are constants $A<\infty$ and $\mu>0$ such that
\begin{equation} \label{eq:expbound}
\int_{\mathcal J} \EE \left( \|\chi_x \frac{1}{H^{(\Lambda_n)}-E}
\chi_y \|^s \right)\,dE \le A e^{-\mu \mbox{\small
dist}\,_{\Lambda_n}(x,y)}
\end{equation}
for all $n\in \NN$ and $x,y \in \Lambda_n$. Then for every
$r<1/(2-s)$ there exists $A_r<\infty$ such that
\begin{equation} \label{eq:dynbound}
\EE \left( \sup_{g:|g|\le 1} \|\chi_x g(H^{(\Omega)}) P_{\mathcal
J}(H^{(\Omega)}) \chi_y \|_{\tr} \right) \le A_r e^{-r\mu
\mbox{\small dist}\,_{\Omega}(x,y)}
\end{equation}
for every $x,y \in \Omega$. Here the supremum is taken over all
Borel measurable functions $g$ which satisfy $|g|\le 1$ pointwise.

In the case $\Omega = \RR^d$ it further holds that the spectrum of
$H$ in $\mathcal J$ is almost surely pure point, with
eigenfunctions $\psi$ which for every $\nu \in (0,2/(2-s))$
satisfy
\begin{equation} \label{eq:efdecay}
\limsup_{|x|\to\infty} \frac{\ln |\psi(x)|}{|x|} \le -\nu \;.
\end{equation}
\end{thm}

The bound (\ref{eq:dynbound}) with $g(H) = e^{itH}$ implies
dynamical localization with exponential decay of the transition
amplitudes for wave packets with energies restricted to $\mathcal
J$. This is stronger than the dynamical bounds which can be
obtained through the multiscale analysis approach, e.g.\
\cite{Germinet/Klein} for the best known result.

Theorem~\ref{thm1} as well as Theorem~\ref{thm2} below are
applicable even when the operator exhibits extended boundary
states in certain geometries, provided there is ``localization in
the bulk''. This is the relevance of the domain adapted metric
$\dist_{\Omega}$. Note that $\dist_{\RR^d}(x,y)=|x-y|$.

While typical applications of Theorem~\ref{thm1} (see Section~5)
will work with exponential bounds for $\EE (\|\chi_x
(H^{(\Lambda_n)}-E)^{-1} \chi_y \|^s)$ which are uniform in $E\in
{\mathcal J}$, it is interesting to note that Theorem~\ref{thm1}
only requires the energy-averaged bound (\ref{eq:expbound}).

The proof of Theorem~\ref{thm1} in \cite{AENSS} proceeds by first
verifying the bound (\ref{eq:dynbound}) for the finite volume
operators $H^{(\Lambda_n)}$, with constants uniform in $n$. In
finite volume the norm of $\chi_x g(H) P_{\mathcal J}(H) \chi_y$
may be estimated in terms of sums of bounds on rank-one operators.
The latter have equal operator and trace norms, which ultimately
allows to state (\ref{eq:dynbound}) as a trace norm bound.

\section{A finite volume criterion}

In applications of Theorem~\ref{thm1} it is necessary to find
energy regimes in which the exponential resolvent bound
(\ref{eq:expbound}) can be verified. In this section we provide a
finite volume sufficiency criterion for the desired exponential
decay.

We define the boundary layer of a set
\begin{equation} \label{eq:boundary}
\delta \Lambda := \{ q: r< \dist(q,\Lambda^c) < 23r \}\;,
\end{equation}
where the choice of the depth is somewhat arbitrary, but
convenient for the technical implementation of the proof of the
following result.

\begin{thm} \label{thm2}
Let $H$ be as above. Then for each $s\in (0,1/3)$ and $\lambda >0$
there exists $M_{s,\lambda} <\infty$, such that if for some $E\in
\RR$ and $L> 24r$,
\begin{eqnarray} \label{eq:finvolcrit}
e^{-\gamma} & := & M_{s,\lambda} (1+1/\lambda)^{2s}
(1+|E-E_0|)^{5s(d+2)} (1+L)^{2(d-1)} \nonumber \\
& & \mbox{} \times \limsup_{\varepsilon\to 0} \sup_{\alpha\in
{\mathcal I}} \EE \left( \| \chi_{\alpha}
\frac{1}{H^{(B_{\alpha}^L)}-E-i\varepsilon} 1_{\delta
B_{\alpha}^L} \|^s \right) \nonumber \\
& < & 1 \;,
\end{eqnarray}
then for any open $\Omega \subset \RR^d$ and any $x,y \in \Omega$
\begin{equation} \label{eq:expbound2}
\limsup_{\varepsilon\to 0} \EE \left( \|\chi_x
\frac{1}{H^{(\Omega)}-E-i\varepsilon} \chi_y \|^s \right) \le
e^{\gamma} A(s,\lambda,E) e^{-\gamma \dist_{\Omega}(x,y)/2L} \;,
\end{equation}
with $A(s,\lambda,E)$ the right hand side of
(\ref{eq:boundedness}). One may choose
\begin{equation} \label{eq:Mbound}
M_{s,\lambda} = \const \frac{(1+\lambda)^{5s(d+4)}}{1-3s} \;.
\end{equation}
\end{thm}

The proof of Theorem~\ref{thm2} proceeds essentially by an
iterative argument where the distance from $x$ to $y$ is covered
by balls of radius $L$. The bound (\ref{eq:finvolcrit}) serves as
an initial decay estimate for the resolvent between the center and
boundary of a ball of radius $L$, reflected in the positive
exponent $\gamma$. An iterative geometric resolvent expansion is
used to show that the decay adds up (or better: multiplies up) to
exponential decay with rate proportional to $\gamma$ and
$\dist_{\Omega}(x,y)$. As this resolvent expansion does not work
near the boundary of $\Omega$, one uses the modified
$\dist_{\Omega}$.

The factors which appear in each step of the resolvent expansion
are not all independent. Therefore a (triple) H\"older bound is
used to factorize their expectations. This is the reason for
having to work with $s<1/3$. In order to not having to divide the
exponent $s$ by three in each step of the iteration (which would
cause it to collapse into $0$), the random parameters
$\eta_{\alpha}$ near the boundaries of domains used in the
expansion are re-sampled in each step of the iteration. This means
that they are replaced with parameters $\hat{\eta}_{\alpha}$ which
are independent of the $\eta_{\alpha}$, but have the same
distribution, a procedure which also appears, for example, in the
spectral averaging argument of \cite{Simon/Wolff}. This allows to
avoid the use of various versions of ``decoupling lemmas'' which
have entered the fractional moment method for lattice models and
seem to be harder to verify in the continuum.

As opposed to the use of an iteratively increasing sequence of
length scales in the multiscale analysis approach, only one length
scale $L$ is used by the fractional moment method to go from
finite to infinite volume. In this iteration process,
Lemma~\ref{lemma} plays a role similar to Wegner estimates in
multiscale analysis. It provides a worst case bound on the growth
of the resolvent over distances less than $L$, where
(\ref{eq:finvolcrit}) can not yet be used.

The exponential decay bound (\ref{eq:expbound2}) on resolvents is
not only a necessary consequence of the finite volume criterion
(\ref{eq:finvolcrit}), but for $\Omega =\RR^d$ is also sufficient
for it, as shown by the following result:

\begin{thm} \label{thm3}
Let $H$ be as above and suppose that for some $A<\infty$, $\mu >0$
and $E\in \RR$
\begin{equation} \label{eq:expbound3}
\limsup_{\varepsilon \downarrow 0} \EE \left( \|\chi_{\alpha}
\frac{1}{H-E-i\varepsilon} \chi_{\beta} \|^s \right) \le Ae^{-\mu
|\alpha -\beta|}
\end{equation}
for all $\alpha, \beta \in \ZZ^d$. Then, for sufficiently large
$L$, (\ref{eq:finvolcrit}) is satisfied uniformly for all $E'$ in
an open neighborhood of $E$.
\end{thm}

A particular consequence of this is that Theorem~\ref{thm1} could
be stated under assuming the infinite volume exponential decay
bound (\ref{eq:expbound3}) for the resolvent, as
(\ref{eq:expbound3}) implies (\ref{eq:finvolcrit}) and
(\ref{eq:finvolcrit}) implies (\ref{eq:expbound2}) (which allows
for finite volume) and thus (\ref{eq:expbound}) for a neighborhood
$\mathcal J$ of $E$ and all $\Lambda_n$.

It is interesting to note that Theorem~\ref{thm3} allows to
conclude localization on an open interval from a bound for a
single energy. This is due to the fact that for finite volume
$\Lambda$ the fractional moments $\EE (\| \chi_x
(H^{(\Lambda)}-z)^{-1} \chi_y \|^s)$ are H\"older continuous in
$z\in \CC$. Thus the set of energies where a bound like
(\ref{eq:finvolcrit}) is valid must be open.

\section{Applications}

Applications of our method consist in verifying the bound
(\ref{eq:finvolcrit}) in concrete energy regimes. In this sense
(\ref{eq:finvolcrit}) is a fractional moment version of the
initial length bounds used to start a multiscale analysis. Here
are examples of regimes where (\ref{eq:finvolcrit}) can be
verified (for detailed statements see \cite{AENSS}):

\begin{itemize}
\item The band edge/Lifshitz tail regime: Here
(\ref{eq:finvolcrit}) follows from smallness of the density of
states in a suitable energy interval. One may work with smallness
of the expected number of eigenvalues of finite volume operators
$H^{(\Lambda)}$ or directly with smallness of the integrated
density of states in infinite volume, e.g.\ Lifshitz tails.

\item The large disorder regime: Under somewhat stronger
assumptions on the distribution of the $\eta_{\alpha}$, for
example in the case of uniform distribution on $[0,1]$, one can
improve the bounds (\ref{eq:Cbound}) and (\ref{eq:Mbound}) and
obtain $C_{s,\lambda}$ and $M_{s,\lambda}$ which are bounded as
$\lambda \to \infty$. This in turn may be used to prove
localization in the large disorder regime: For every $E'\in \RR$
there exists $\lambda'$ sufficiently large such that for $\lambda
> \lambda'$ the energy interval $(-\infty,E')$ is localized (i.e.\
(\ref{eq:finvolcrit}) can be verified for all $E\in
(-\infty,E')$).

\item The multiscale analysis regime: One may also use the typical
output of a multiscale analysis to verify (\ref{eq:finvolcrit}) at
sufficiently large $L$. Thus one gets the stronger dynamical
localization bounds provided by the fractional moment method in
all regimes where a multiscale analysis can be carried through and
our general setup from Section~2 holds.
\end{itemize}

\section*{Acknowledgements}
 In the course of this project, the authors' work was supported in part by
 NSF grants PHY-9971149 (MA, and AE),  DMS 0070343 and
 0245210 (GS), INT-0204308 (JS),
 NSF postdoctoral research fellowship (JS), and a NATO
 Collaborative Linkage Grant PST.CLG.976441 (GS and SN).
 The authors also thank Caltech, Universit\'{e} Paris 7 and
 the Institute Mittag-Leffler for hospitality which has facilitated
 this collaboration.


\vspace{.5cm}

\noindent {\bf Author affiliations and e-mail:}

M.~Aizenman, Princeton University, aizenman@princeton.edu

A.~Elgart, Stanford University, elgart@CIMS.nyu.edu

S.~Naboko, St.~Petersburg State University, naboko@math.su.se

J.~H.~Schenker, ETH Z\"urich, jschenker@itp.phys.ethz.ch

G.~Stolz, University of Alabama at Birmingham, stolz@math.uab.edu

\end{document}